\documentclass[%
reprint,
superscriptaddress,
amsmath,amssymb,
aps,
floatfix,
]{revtex4-2}
\usepackage{float}
\usepackage{graphicx}
\usepackage{dcolumn}
\usepackage{bm}
\usepackage{balance}
\usepackage{graphicx}
\usepackage{dcolumn}
\usepackage{siunitx}
\usepackage{xr}
\usepackage{xcolor}
\usepackage[utf8x]{inputenc}
\usepackage[normalem]{ulem}
\makeatletter
\usepackage{microtype}
\begin{document}


\author{S. Arora}
\author{T. Bauer}
\affiliation{Kavli Institute of Nanoscience, Delft University of Technology, 2600 GA, Delft, The Netherlands}
\author{R. Barczyk}
\author{E. Verhagen}
\affiliation{Center for Nanophotonics, AMOLF, Science Park 104, 1098 XG Amsterdam, The Netherlands}
\author{L. Kuipers}\email{l.kuipers@tudelft.nl}
\affiliation{Kavli Institute of Nanoscience, Delft University of Technology, 2600 GA, Delft, The Netherlands}

\title{Multiple backscattering in trivial and non-trivial topological photonic crystal edge states with controlled disorder}

\date{\today}

\begin{abstract}

We present an experimental investigation of multiple scattering in photonic-crystal-based topological edge states with and without engineered random disorder. We map the spatial distribution of light as it propagates along a so-called bearded interface between two valley photonic crystals which supports both trivial and non-trivial edge states. As the light slows down and/or the disorder increases, we observe the photonic manifestation of Anderson localization, illustrated by the appearance of localized high-intensity field distributions. We extract the backscattering mean free path (BMFP) as a function of frequency, and thereby group velocity, for a range of geometrically engineered random disorders of different types. For relatively high group velocities (with $n_g < 15$), we observe that the BMFP is an order of magnitude higher for the non-trivial edge state than for the trivial. However, the BMFP for the non-trivial mode decreases rapidly with increasing disorder. As the light slows down the BMFP for the trivial state decreases as expected, but the BMFP for the topological state exhibits a non-conventional dependence on the group velocity. Due to the particular dispersion of the topologically non-trivial mode, a range of frequencies exist where two distinct states can have the same group index but exhibit a different BMFP. While the topological mode is not immune to backscattering at disorder that breaks the protecting crystalline symmetry, it displays a larger robustness than the trivial mode for a specific range of parameters in the same structure. Intriguingly, the topologically non-trivial edge state appears to break the conventional relationship between slowdown and the amount of backscattering.

\end{abstract}
\maketitle

\section{Introduction}
Robust scattering-free photonic transport is a major driving force in device miniaturization, on-chip photonic networks, and photonic quantum technology \cite{butt2021recent, bogaerts2020programmable,yu2021ultra, chen2021highlighting}. In principle, using slow light will benefit developments through the concomitant increase in light-matter interactions \cite{baba2008slow, krauss2007slow, notomi2001extremely}. However, the much-wanted increased light-matter interaction itself has unwanted yet fundamental side effects. No real-life photonic structure is perfect and as the light slows down, imperfections increasingly scatter the propagating light \cite{smith2000low, hughes2005extrinsic}. Moreover, the fraction of light being backscattered will grow with increased slowdown, inevitably resulting in Anderson localization for one-dimensional systems like waveguides \cite{PhysRevLett.99.253901, sapienza2010cavity}. The advent of crystalline topological insulators in electronic as well as photonic and acoustic systems spurred investigations into robust transmission channels that are inherently protected from scattering \cite{hasan2010colloquium, ozawa2019topological}. These systems offer the possibility of topological transport without a magnetic field that breaks time-reversal symmetry. Instead, the protection is directly linked to crystalline symmetry. The realization of photonic analogs consisting of topologically non-trivial photonic crystals (PhCs) \cite{Wu2015, mittal2019photonic, barik2018topological, Parappurath2020, khanikaev2017, hafezi2013imaging, ozawa2019topological, khanikaev2013photonic, raghu2008analogs, haldane2008possible} with reduced backscattering \cite{yang2013experimental, lu2017observation, wang2009observation, poo2011experimental, minkov2018unidirectional} offer passive implementation and ability for on-chip, nanoscale integration. In the specific case of quantum valley Hall effect emulating systems \cite{Ma2016, Shalaev2018, He2019, arora2021direct,tang2022topological} that operate below the light line, robustness against backscattering relies fully on geometric symmetries.This immediately raises the question to what extent the topological protection holds in the presence of random disorder, which by its very nature destroys the underlying and required symmetry \cite{Orazbayev2019, Arregui2021}.

To address this intriguing question, one must simultaneously consider photonic slowdown and topological robustness. As the propagating light slows down, light becomes more susceptible to backscattering due to a combination of an increase in light-matter interaction and local density of states (proportional to the slow down) that increases the overall scattering at imperfections, and increases the fraction of scattered light into the backward direction which is also proportional to the density of states of the mode and the slow down factor
\cite{hughes2005extrinsic, o2007dependence, andreani2007light}. Recently, a design called bearded interface has been introduced that offers both a topologically trivial and non-trivial mode within the bandgap of the same silicon-on-insulator device, i.e., with the same unavoidable fabrication-induced disorder \cite{yoshimi2020slow}. Moreover, this interface offers the prospect of achieving two topologically distinct modes over an extended range of achievable group velocities. Initial transmission measurements have found that the transmission along straight and symmetry-preserving corners of the non-trivial mode was indeed higher than for the trivial mode \cite{yoshimi2020slow, Yoshimi2021, JalaliMehrabad2020}, with losses increasing for slower light. A more recent study on scattering losses due to intrinsic fabrication disorder found no appreciable difference between the robustness of topologically trivial and non-trivial edge states \cite{rosiek2022observation} with all losses being enhanced by the slowdown of the propagating light. By observing indications of the onset of Anderson localization, the study suggested the presence of multiple (back)scattering. In all cases, the determination of the group velocity came from simulations. It is important to note that, interestingly, due to small differences in the exact geometrical parameters (relative hole sizes and corner rounding) of the realized photonic crystals, the frequency for which the group velocity is lowest is not equivalent in both investigations: in one it is located inside the non-trivial frequency range \cite{Yoshimi2021} and in the other, inside the trivial frequency range \cite{rosiek2022observation}. This highlights, that for probing the limits of topological protection, investigations that use engineered disorder, combined with a local determination of the group velocity would be highly valuable.

In this paper, we experimentally quantify the robustness of edge states propagating along a bearded interface using a universal metric known as backscattering mean free path (BMFP) \cite{PhysRev.109.1492, shi2015statistics, schwartz2007transport}. We directly map the spatial distribution of the complex electric field of the edge states with phase-resolved near-field scanning optical microscopy \cite{Rotenberg2014, gersen2005direct}. We first investigate the bearded interface without induced disorder and systematically compare the BMFP of the trivial and non-trivial modes for the same range of group velocities. Subsequently, we study the effect of intentionally induced positional or hole size disorders of various magnitudes. While both trivial and non-trivial waveguides show scattering losses, we observe a different qualitative and quantitative response for the topologically trivial and non-trivial edge states, which depends on the type and magnitude of the disorder. For the largest magnitudes of the disorder, we observe the photonic manifestation of Anderson localization, illustrated by the appearance of localized high-intensity states. Interestingly, we find that the onset of Anderson localization is different for the distinct edge states. We also observe that the conventional direct relation between the slowdown factor and backscattering seemingly breaks down for the topologically non-trivial mode.

\section{Experimental Results}

\subsection{Mapping of the near-field edge state}

We experimentally investigate edge states at the interface of topological PhCs mimicking the quantum valley Hall effect. A non-vanishing Berry curvature at the valleys is associated with an intrinsic magnetic moment, resulting in a valley Chern number $C_{K, K^{\prime}} = \pm \frac{1}{2}$ \cite{shalaev2019robust, cheng2016robust, Xiao2007, zak1989berry, Ma2016, Hafezi2011, xie2018photonics, Kim2014, Dubrovkin2020}. This gives rise to photonic bulk bands of a topologically non-trivial nature. For a unit cell with underlying $C_3$ symmetry, there are several ways to realize valley-protected edge state interfaces, akin to edge terminations in a honeycomb lattice structure such as graphene. Edge state modes appear at the K and K$^\prime$ points of the Brillouin zone (BZ) as a consequence of broken lattice symmetry \cite{Tzuhsuan2016, Hafezi2011, xie2018photonics, kim2014topological, dubrovkin2020near} by mirror inversion of the PhC lattices at the interface where $y=0$. The two valley-protected degenerate eigenmodes traverse the photonic band gap with a linear dispersion. This is referred to as a zigzag interface \cite{shalaev2019robust, cheng2016robust}. The bearded interface combines parity inversion and glide symmetry \cite{yoshimi2020slow} and the resulting interface  enforces a degeneracy at the BZ edge \cite{mock2010space} resulting in two eigenmodes in each valley with a slow-light region around the BZ edge. The high-energy eigenmode is attributed to have a trivial nature as it can be observed to `drop down' from the conduction band when holes at the interface are adiabatically reduced in size \cite{yoshimi2020slow}. At the same time, the low-energy eigenmode is attributed to be non-trivial. It is observed to exhibit higher transmission through sharp bends in comparison to the high-energy eigenmode, as is expected for a non-trivial edge state \cite{Yoshimi2021}. From this point on, the low-energy mode is thus referred to as the non-trivial mode, whereas the high-energy mode is called the trivial mode. The continuous deformation of the interface holes and the resulting glide plane symmetry gives rise to the two eigenmodes bending towards the BZ edge. This results in a slow-light region for both the trivial and non-trivial modes enabling access to different backscattering regimes \cite{le2009light}.

\begin{figure}[hbt]
	\centering
    \includegraphics[width=\linewidth]{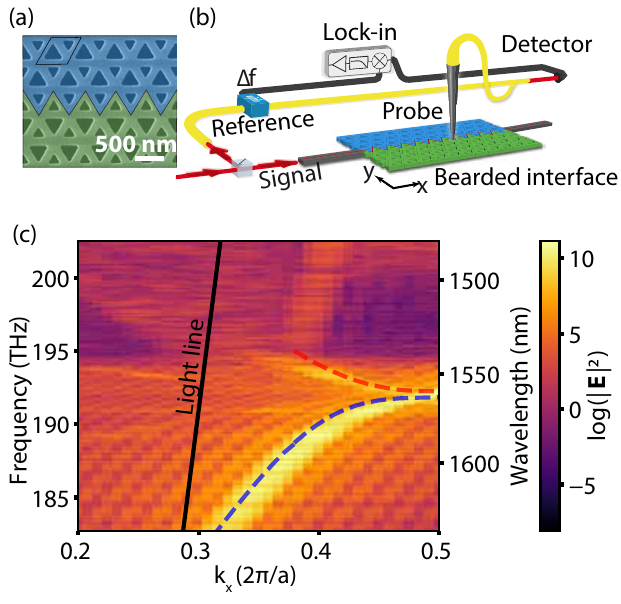}
	\caption{(a) Scanning electron micrograph (SEM) of the bearded interface in the fabricated sample with the color-coded regions depicting the two mirror-inverted lattices with glide symmetry (blue and green). The lattice periodicity is $a=\SI{510}{nm}$. (b) Schematic representation of the experimental setup employed to study the near-field of the bearded interface with amplitude and phase resolution. (c) The experimentally measured dispersion diagram of the first Brillouin zone of the edge state in a pristine PhC with two edge modes: at lower frequencies, non-trivial and at higher frequencies, a trivial mode that is degenerate at \SI{1558}{nm} or \SI{192.4}{THz}. The blue and red dashed lines indicate the numerically calculated dispersion curves of the non-trivial and trivial modes, respectively.}
	\label{ch6-fig1}
\end{figure}

\begin{figure*}[hbt]
	\centering
    \includegraphics[width=\linewidth]{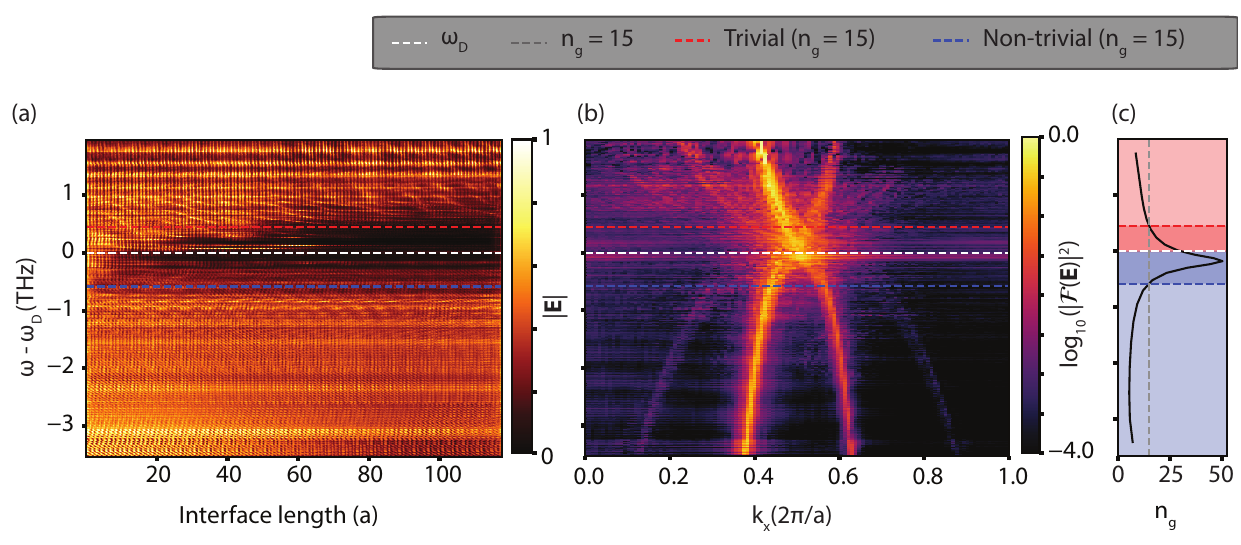}
	\caption{Real and reciprocal space of electromagnetic field of an ensemble-averaged realization of a bearded interface without the intentional disorder. (a) Near-field spatial-spectral amplitude map normalized to the incoupling region collected as a function of frequency and position along the interface. Light propagates the TPC from the left. (b) Ensemble-averaged dispersion curve.  The white dashed horizontal line denotes where trivial and non-trivial modes become degenerate at the Brillouin zone edge. (c) Calculated group index curve from the measured ensemble averaged dispersion. The blue and red shaded regions depict the non-trivial and trivial modes separated by the degeneracy point given by the white dashed line at $\SI{0}{THz}$. The dark blue and dark red regions indicate the slow light region for the non-trivial and trivial modes with a group index of $n_g > 15$,  respectively.}
	\label{ch6-fig2}
\end{figure*}

We fabricate the bearded interface in a silicon-on-insulator platform (pseudo-colored in SEM image Fig. \ref{ch6-fig1}(a)). The unit cell with lattice constant $a=\SI{510}{nm}$ consists of equilateral  triangular holes with side lengths $L_1 = 0.6a$ and $L_2 = 0.4a$. An interface is created by transforming the triangular holes at the interface of one lattice from $L_1$ to $L_2$. We measure the complex in-plane optical fields propagating along the length of the PhC spanning $180$ unit cells as a function of laser excitation wavelength $\lambda = [\SI{1480}{nm} - \SI{1640}{nm}]$  using phase-sensitive near-field scanning optical microscopy \cite{Rotenberg2014}. The employed heterodyne detection scheme allows us to decompose each spatial field map into its Fourier components $\mathcal{F}(k_x, k_y)$. Thus, we obtain the mode dispersion. Figure \ref{ch6-fig1}(b) shows the dispersion of a bearded interface without any engineered disorder as an example. The region above $\SI{194.5}{THz}$ represents the start of top bulk bands. Rather than a linear dispersion as supported by the zigzag interface \cite{arora2021direct}, we observe both predicted modes each with a steep linear slope around $k_x =  2\pi/3a$, and also a slow light region close to the Brillouin zone edge ($k_x = \pi/a$). The red (trivial) and blue (non-trivial) dashed outlines indicate an excellent correspondence with numerically simulated dispersion curves for the two modes.

\subsection{Visualizing real and reciprocal space of the edge state}

We measure the electric field of light that enters the crystal through a feed waveguide and propagates along the interface until it leaves at the end-facet, which serves as a termination of the PhC. An exemplary measured electric field amplitude of a bearded interface without engineered disorder is shown in Supplementary Fig. S1. The end-facet in addition to fabrication imperfections such as surface and side wall roughness may cause backscattering and create intrinsic loss channels \cite{arora2021direct,melati2014real}. The procedure that we follow for all spatial-spectral maps in this work includes obtaining an ensemble average of five realizations of bearded interface PhCs with the same magnitude of disorder, summed over modes with the same group velocity \cite{topolancik2007random}. Due to slight variations in nanofabrication, the dispersion relation of individual PhCs may exhibit slight shifts in the degeneracy point $\omega_D$  (DP) of the trivial and non-trivial modes. We therefore shift all measurements to ensure that the DPs coincide before averaging. As a consequence, the vertical axis of Fig. \ref{ch6-fig2} indicates modified units of frequency in terahertz denoting the frequency difference with respect to that of the DP ($\omega - \omega_D$). The spatial-spectral amplitude map displayed in Fig. \ref{ch6-fig2}(a) shows the ensemble-averaged electric field amplitude as a function of  $\omega - \omega_D$ and interface length (in units of lattice constant $a$). To account for the excitation-dependent incoupling efficiencies, each horizontal line in the spatial-spectral map is normalized to the mean amplitude of the first four unit cells after the feed waveguide. The corresponding ensemble-averaged dispersion is shown in Fig. \ref{ch6-fig2}(b). The white dashed line denotes the DP of the trivial and non-trivial mode in the ensemble dispersion at the boundary of the Brillouin zone edge. By fitting a hexic polynomial function (polynomial of degree six) to the measured ensemble-averaged dispersion curve we determine the group velocity as a function of frequency. The extracted group index for the ensemble $n_g = c/v_g$, where $c$ is the speed of light in vacuum and $v_g$ is the group velocity of the measured edge state, is shown in Fig. \ref{ch6-fig2}(c). More information on this procedure is provided in Supplementary Sec. III. 

For qualitative scrutiny of the ensemble-averaged spatial-spectral map (Fig. \ref{ch6-fig2}(a)), we separate the edge mode of the bearded interface into two regimes: fast dispersive ($n_g \le 15$) and slow diffusive ($n_g > 15$). Below \SI{-0.66}{THz} (blue dashed line), the non-trivial mode has a group index of $n_g \le 15$. Here, the non-trivial mode retains a nearly uniform amplitude distribution over the entire interface length consistent with unity transmission \cite{arora2021direct}, and the corresponding dispersion shows a positive steep slope within $k_x < \pi/a$.  Amplitude oscillations (with a spatial frequency of roughly $2\pi/3a$ and lower) in the spatial-spectral maps arise due to the interference of the backward and forward propagating Bloch modes. As the non-trivial mode enters the diffusive regime ($n_g > 15$) for larger frequencies, the mode nominally slows down further and the amplitude starts to rapidly decay with distance. In reciprocal space (Fig. \ref{ch6-fig2}(b)) more spatial frequencies appear around the DP. We observe that the non-trivial mode reaches a maximum group index of $n_g = 50$ at $\omega - \omega_D = -0.17$ (shown in Fig. \ref{ch6-fig2}(c)). Importantly, this maximum group index is found to occur not at the DP but at a lower frequency, i.e., in the frequency range of the non-trivial edge state (see Supplementary Sec. V and ref. \cite{rosiek2022observation}). As we cross over to the trivial region, we observe a gradual decrease of $n_g$ as the frequency is increased away from the BZ edge with a group index $n_g = 30$ at the BZ edge. The influence of unavoidable fabrication-induced disorder on the trivial mode is evident. A clear indication of light undergoing backscattering along the waveguide is visible in the frequency region around the DP. The local high-intensity features are characteristic of Anderson localized modes \cite{Smolka2011, spasenovic2012measuring}. Above the red dashed line, where the trivial mode has $n_g \le 15$, the dispersion is roughly linear. However, multiple interference patterns along the interface are observed in the spatial-spectral maps with a distinct decay, in sharp contrast to the homogeneous distribution of the non-trivial mode with the same $n_g$ range. Thus, from qualitative scrutiny, it is apparent that the trivial edge state suffers more backscattering than the non-trivial edge state. 

\begin{figure}[htbp!]
	\centering
    \includegraphics[width=\linewidth]{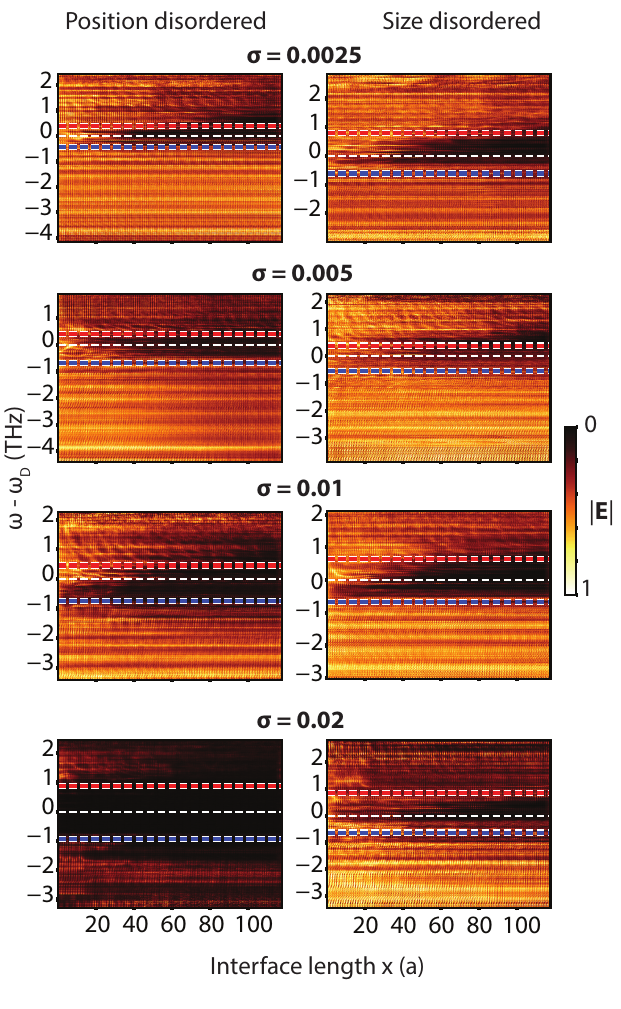}
	\caption{Spatial-spectral maps of the measured electrical field amplitude (normalized to the input waveguide) for various degrees of intentionally induced disorder averaged over five realizations of the disorder. The two columns represent position- and size-induced disorder, respectively, as a function of interface length in units of lattice constant $a=\SI{510}{nm}$. The white dashed line denotes DP of trivial and non-trivial mode. The red and blue dashed lines (with a white background for better visibility) denote the frequency away from DP where the trivial and non-trivial mode has a group index of $n_g = 15$, respectively.}
	\label{ch6-fig3}
\end{figure}

\subsection{Introducing random intentional disorder}

To experimentally unravel the degree of protection that a trivial and non-trivial mode may offer against multiple scattering, we intentionally introduce disorder in the bearded interface PhC design. We engineer two types of intentional disorder: (a) random displacement of the holes and (b) hole size variations. We introduced several magnitudes of disorder for each type of disorder and fabricated five different PhC interface realizations per magnitude. The triangular holes are randomly displaced and deformed in accordance with a normal distribution with standard deviation $\sigma$ as a measure of position and size disorder. The value of $\sigma$ is scaled with the lattice constant $a$ and the hole diameters $L_1$ and $L_2$ (which are also in units of $a$) for position disorder and size disorder, respectively. Figure \ref{ch6-fig3} shows the ensemble-averaged spatial-spectral amplitude maps for position and size disorders with increasing disorder magnitude $\sigma = [0.0025, 0.005, 0.01, 0.02]$. To account for variations in in-coupling efficiency for different excitation frequencies, each horizontal line on the spatial-spectral maps is also normalized to the mean intensity at the feed waveguide. Qualitatively, the following observations can be made for the non-trivial (bottom) and trivial (top) modes as we increase the magnitude of the engineered disorder. Every ensemble-averaged spatial-spectral map in Fig. \ref{ch6-fig3} reveals that the light in the fast non-trivial mode region (the limit denoted by the blue dashed line where $n_g = 15$) exhibits an almost homogeneous spatial profile. Interference of the forward and backward modes due to back reflection at the end is evidenced by fast amplitude oscillations. As the frequency is increased into the slow non-trivial region close to the DP, the intensity along the interface rapidly decays to a negligible value. The propagation losses become prevalent and the penetration of the light along the interface decreases. Similar behavior is evident in the slow trivial region below the red dashed line where $n_g = 15$. However, the fast trivial region above the red dashed line looks different than that of the non-trivial mode, showing multiple interference patterns along the interface. In the vicinity of the DP, we observe high-intensity resonances that feature spectral and spatial dependence. These sharp features of intensities are fingerprints of localized states and they become more pronounced as the magnitude of the disorder increases. In general, we observe a rather sharp transition in propagation behavior for the non-trivial region as the mode slows down as the frequency increases, from an edge state that propagates along the entire interface to a state that is hardly able to penetrate the structure. In contrast, the transition in propagation behavior from fast to slow light in the trivial mode is more gradual  with frequency. We scale the magnitude of both types of engineered disorder to the lattice constant in a linear fashion. If we use a similar scaling as was used in previous works for both two-dimensional and three-dimensional photonic crystals \cite{extinction2005, wiersma2013disordered, Orazbayev2019} to compare the effects of different types of disorder, it suggests that the influence of position disorder is more prominent on edge state propagation than size disorder, as evidenced by the slower onset of speckle-like interference patterns in trivial and non-trivial regions. 

\section{Quantifying backscattering}

To quantify the robustness to backscattering of the non-trivial mode in comparison to the trivial mode in the bearded interface, we use the backscattering mean free path (BMFP or $\xi$) as a figure of merit \cite{arcari2014near, garcia2010density}. It is defined as the absolute displacement along the bearded interface after which light suffers (multiple) scattering events in the absence of other loss channels such as absorption \cite{skipetrov2016red} or out-of-plane losses. 

\begin{figure}[hbt]
	\centering
    \includegraphics[width=0.5\textwidth]{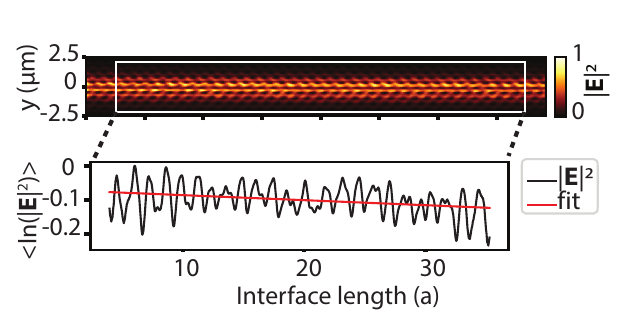}
	\caption{(a) Normalized ensemble-averaged electromagnetic field intensity map, averaged over 5 realizations of bearded interface without intentionally induced disorder of the non-trivial mode with integrated over fast light region. Inset: Ensemble-averaged field intensity profile fitted with a linear slope to obtain the backscattering mean free path. }
	\label{ch6-fig4}
\end{figure}

We extract the backscattering mean free path along the interface for an exemplary waveguide as shown in the inset of Fig. \ref{ch6-fig4} using the following relation \cite{beer1852bestimmung}: 
\begin{equation}\label{eq:1}
    - \frac{x}{\xi(\omega)}  = \left\langle\ln\left[\frac{I(\omega)}{I_0(\omega)}\right]\right\rangle 
\end{equation} where $x$ signifies the distance from the feed waveguide, $\frac{I}{I_0}$ is the normalized electric field intensity, and the brackets indicate the statistical ensemble average over different PhCs with the same type and magnitude of disorder. The frequency ($\omega$) dependence of $\xi$ is stated to differentiate between the contributions of disorder on the non-trivial and trivial modes. 

\begin{figure}[hbt]
	\centering
    \includegraphics[width=\linewidth]{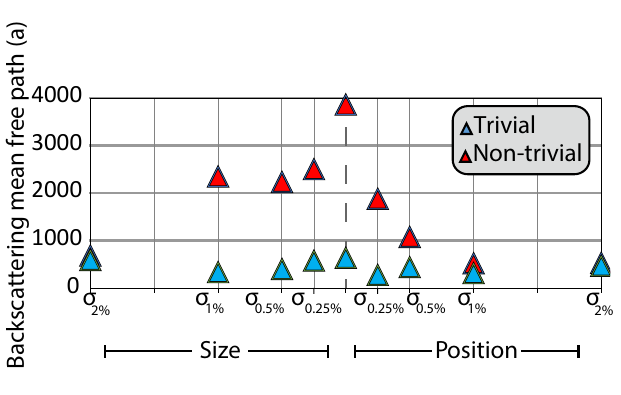}
	\caption{Measured backscattering mean free path as a function of intentionally induced disorder. The triangles indicate the measured BMFP from each ensemble averaged disorder realization for the fast non-trivial (red) and fast trivial (blue) modes. The left and right side of the graph represents the fabricated size and position-induced disorder respectively. }
	\label{ch6-fig5}
 
\end{figure}
\begin{figure*}[hbt]
	\centering
    \includegraphics[width= \textwidth]{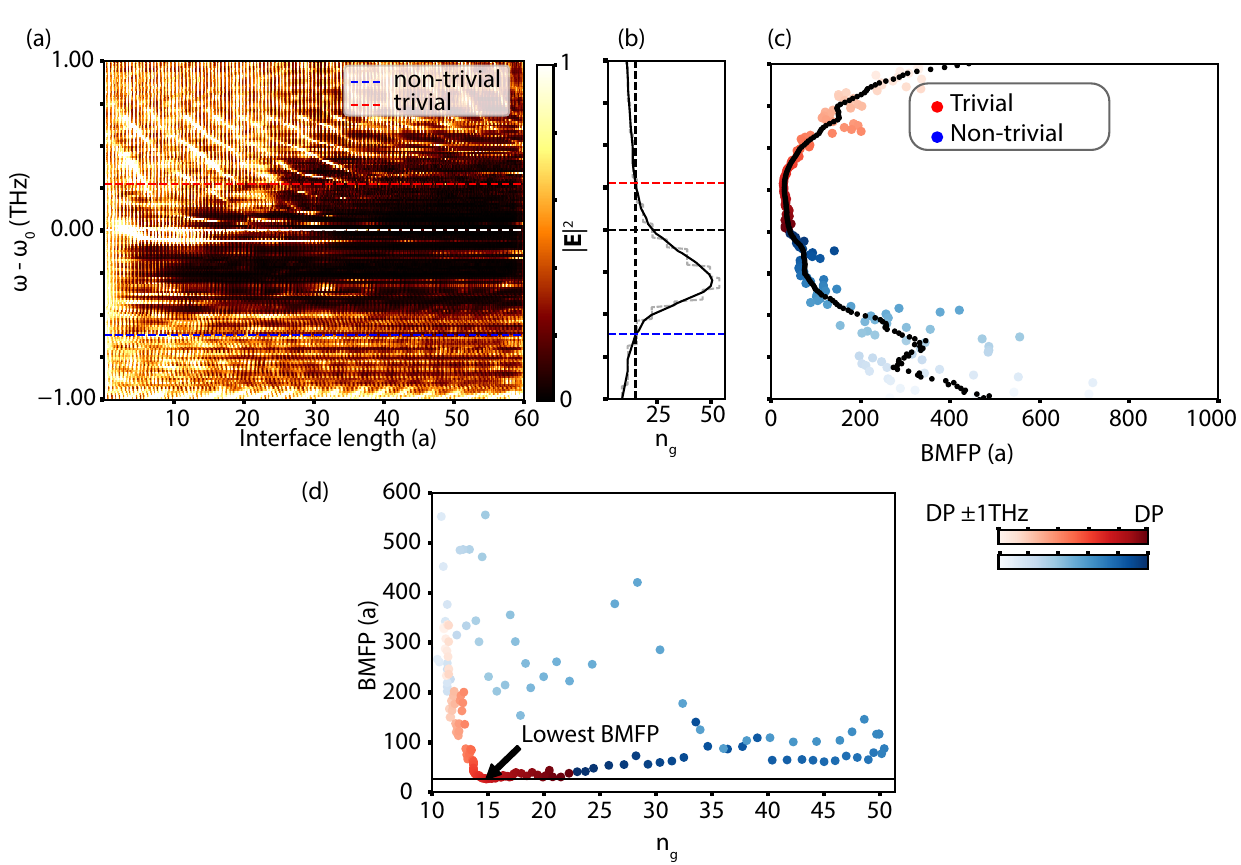}
	\caption{Qualitative analysis of the slow light region. a) Zoom-in of Fig. \ref{ch6-fig2}(a) denoting the spatial-spectral map confined to the slow light region of the two modes. b) Group index curve for the frequencies in the range $[-1, 1]$, with the red and blue dashed line denoting the frequencies where $n_g = 15$ for the trivial and non-trivial region, respectively. c) Backscattering mean free path as a function of frequency away from DP. The red and blue points denote the measured BMFP, whereas the black curve denotes the non-trivial and trivial BMFP for a sliding window of width \SI{0.17}{THz}. It is clear that the minimum BMFP neither occurs at the DP nor the region of the maximum group index. d) Backscattering mean free path as a function of $n_g$. The blue and red circles denote the non-trivial and trivial mode for a sliding window of width \SI{0.17}{THz}, respectively. The increasing darkness of the filled circles indicates the relative frequency away from the DP where the darkest red and darkest blue denote the slow light region, whereas light red and light blue denote fast light of trivial and non-trivial mode, respectively.}
	\label{ch6-fig6}
\end{figure*}

The experimentally determined backscattering mean free path $\xi$ as a function of induced disorder is plotted in Fig.  \ref{ch6-fig5}. The red and blue triangles indicate the BMFP of fast dispersive regimes ($n_g \le 15$) for trivial and non-trivial modes in the ensemble  with an equal group index range, respectively (see Supplementary Table. I for BMFP values and mean group indices of the trivial and non-trivial regions). For the PhC ensemble with no intentionally engineered disorder ($\sigma_{0}$), the non-trivial mode is more robust to backscattering by an order of magnitude compared to the trivial mode. As the magnitude of the disorder increases, the BMFP for the non-trivial mode drops by almost an order of magnitude for the largest disorder. The decrease in BMFP is strongest for increases in positional disorder. Surprisingly, the trivial mode doesn't suffer a similar reduction and exhibits a BMFP that remains roughly constant with increasing disorder \cite{Orazbayev2019}. Importantly, we observe backscattering for all degrees of engineered disorder, and thus symmetry-induced protection partly breaks down due to the induced disorder. However, the non-trivial mode suffers less backscattering than the trivial mode of the waveguide without intentional disorder. This suggests that a certain degree of robustness of the non-trivial mode remains in effect for the studied types and magnitudes of disorder. Nevertheless, the rapid decrease in BMFP does suggest a rapid deterioration of the symmetry-induced protection \cite{Arregui2021}. At the maximum value of induced disorder, for size and position ($2\%$), the magnitude of the BMFP for the non-trivial and trivial mode is almost the same, suggesting that for the largest disorder, topological protection is destroyed to a certain degree. It is important to note that for the fast non-trivial and trivial mode without the engineered disorder, the deviation in the mean group indices of the two modes is within $1.05\%$ (where $n_g \le 15$), see Table S1 for BMFP values and mean group indices of the trivial and non-trivial regions of the ensemble. An upper limit of the deviation in the mean group indices of the trivial and non-trivial regions is within $13.39\%$ and it occurs for size disorder $0.5\%$ and does not directly dictate the trend in BMFP. The difference in BMFP observed in Fig. \ref{ch6-fig5} of the two fast modes can therefore not be explained by differences in group velocity. The anomalous value of a low BMFP for both non-trivial and trivial mode at size disorder $\sigma_{0.25\%}$ is attributed to limited statistics. Our observation is in agreement with the work in \cite{Orazbayev2019}, where the initial introduction of disorder results in a quick reduction of BMFP, and thereafter the decay is slower with an increasing amount of disorder. 

\subsection{Group index contribution to backscattering}

A complete understanding of the edge mode's robustness also requires a detailed analysis of the extracted BMFP with respect to light slowing down as defined by the group index $n_g$. In Fig. \ref{ch6-fig6}, we show how the two modes' BMFP depend on  $n_g$. For conventional waveguides, $\xi$ is directly linked to the group index as $ \xi \propto n_g^{-2}$ for ordinary light propagation \cite{hughes2005extrinsic, engelen2008two}. While the concept of a waveguide mode's group index falters in the slow-light regime due to multiple scattering and the diffusive nature of light \cite{le2009light, garcia2010density}, the nominal $n_g$ is typically still useful to describe the average behavior of the light propagation \cite{krauss2007slow, Liam2010}. From Fig. \ref{ch6-fig6}(a) we qualitatively investigate the slow diffusive behavior of the two distinct modes for the case without intentional disorder using the ensemble-averaged spatial-spectral mode profile. Around the DP (white dashed line in Fig. \ref{ch6-fig6}(a) and black dashed line in Fig. \ref{ch6-fig6}(b)), the propagation losses are significant as evidenced by the small penetration along the interface. The non-trivial mode above the blue dashed line where $n_g>15$ suffers a discrete drop in intensity as a function of frequency, nonetheless maintaining a homogeneous distribution along the interface (See Supplementary Fig. S6 for the central spatial-spectral map). In contrast, a gradual decay is observed in the trivial mode as we move from the fast to the slow light regime ($n_g>15$). Figures \ref{ch6-fig6}(c) and Fig. \ref{ch6-fig6}(d) show the experimentally determined BMFP as a function of $\omega-\omega_D$ and $n_g$ for the ensemble without the engineered disorder. It is important to note that the occasional occurrence of localization events for a single frequency along the length of the waveguide can result in meaningless values of the BMFP. To reduce the effect of such occurrences on our investigation of the role of slowdown on the BMFP, we use a sliding averaging window with a width of \SI{0.17}{THz} to obtain a smooth curve denoted by the black circles in Fig. \ref{ch6-fig6}(c). For BMFP as a function of $n_g$ in Fig. \ref{ch6-fig6}(d), the smoothed values in the non-trivial region are denoted as blue circles whereas the smoothed values in the trivial region are depicted in red. The increase in shade of the circles indicates the frequency with respect to $\omega_D$ and therefore also indicates a transition from the fast to the slow light region, with the darkest shade of red and blue denoting the value at the DP. For a narrow range of $10<n_g<15$, the fast non-trivial mode has a higher BMFP than the fast trivial mode. It is interesting to note that the lowest BMFP ($\xi =26.71a$) does not occur at the frequency of the largest group index in the non-trivial region below the DP but occurs when the edge mode lies in the trivial region with $n_g = 15$. When the frequency of the trivial mode decreases further and its slowdown increases as the DP is approached, the BMFP stays roughly constant, if anything increasing slightly. For frequencies lower than the DP where the modes become non-trivial, a small increase of the BMFP is actually observed as the non-trivial modes slow down further towards the maximal slowdown. At the maximal slowdown of $n_g = 50,$ we observe modes whose fields extend through the entire interface with a relatively high BMFP suggesting a degree of robustness against backscattering, a factor of $3.3$ larger than that of light in the comparatively faster trivial mode at $n_g = 15$. Beyond the maximal slowdown peak, the group velocity of the non-trivial mode increases again. For these frequencies, the non-trivial modes will have the same $n_g$ as other modes, either trivial or non-trivial, at frequencies away from the DP. Three interesting observations can be made for these non-trivial modes at frequencies lower than the maximal slowdown. First, the conventional relation between slowdown and degree of backscattering holds again. Second, the BMFP of the non-trivial modes decreases slower with increasing group index than that of the trivial modes. Third, the BMFP of all these modes is always higher than that of the modes above and around the DP with the same $n_g$, regardless of whether the latter is trivial or non-trivial. We suggest that part of the observed differences in BMFP for modes with the same slowdown could be attributed to an effect of the wavevector change on backscattering: the higher the required wavevector change, the smaller the backscattering. This could potentially explain differences in BMFP of non-trivial modes with the same group index and requires further investigation.

Acknowledging the possible misinterpretation when comparing Fig. \ref{ch6-fig6}(a) and Fig. \ref{ch6-fig6}(c), we address the apparent contradiction between the spatial-spectral map and the measured BMFP. When quantifying the BMFP (shown in Fig. \ref{ch6-fig6}(c)) we find that the minimum BMFP occurs for frequencies just below the red dashed line around $\SI{0.39}{THz}$. A cursory glance at the corresponding spatial-spectral map (Fig. \ref{ch6-fig6}(a)) seems to show a large penetration up to $\approx 55a$ for these frequencies. Conversely, frequencies around $\SI{-0.08}{THz}$, directly below the DP, seemingly exhibit the smallest penetration in the spatial-spectral map but show a comparatively high BMFP of $\xi =70.14a$. However, the cursory, visual interpretation of a small penetration in the non-trivial region neglects a much longer intensity tail to the end termination of the PhC. In contrast, the trivial mode around $\omega - \omega_D = \SI{0.39}{THz}$ experiences a drastic reduction in intensity right after the first high-intensity region, even leading to a negligible intensity of light reaching the end termination. This is evident from the spatial-spectral map in the logarithmic intensity scale overlaid on the spatial-spectral amplitude map in Supplementary Fig. S6. These findings emphasize the significance of employing appropriate normalization and background correction methods when assessing the magnitude of and mechanisms behind backscattering.

In conclusion, a direct experimental evaluation of backscattering of fast and slow light propagating along a bearded interface between two topological photonic crystals emulating the quantum valley Hall effect has been performed. The spectral and spatial maps of the full interface enable us to completely visualize and probe the appearance of localized events caused by fabrication disorders. Induced disorder in terms of position and size variations affects the electric field distribution in different ways. For the slowest light, we find an indication of Anderson localization at increasing magnitude of disorder illustrated by the localized areas of high light intensity for both the non-trivial and trivial edge states. However, the sensitivity to disorder is found to be different for the two modes for the investigated range of disorder, with the non-trivial mode exhibiting more robustness than the trivial ones. This study sheds light on the potential relevance of topological crystals for on-chip photonic devices for practical applications. We find that all edge states undergo backscattering, regardless of whether they are trivial or non-trivial. Having said that, our observations do indicate that for the same slowdown, topologically non-trivial edge states can exhibit less backscattering than their trivial counterparts. However, the difference is less than an order of magnitude. Both types of edge states will exhibit multiple backscattering as the light is slowed down significantly or when the amount of disorder increases, ultimately leading to Anderson localization.
\section*{Acknowledgments}

The authors would like to thank S{\o}ren Stobbe, Guillermo Arregui, and Christian Anker Rosiek for bringing to our attention a previous error in the analysis of the BMFP as a function of group index. Their expertise and attention to detail greatly improved the quality of this paper. This work is part of the research programme of the Netherlands Organisation for Scientific Research (NWO). The authors acknowledge support from the European Research Council (ERC) Advanced Investigator Grant no. 340438-CONSTANS and ERC Starting Grant no. 759644-TOPP.

\onecolumngrid
\newpage
\section{Appendix: Methods}

\textit{Measurement technique}
Numerical simulations were performed using MIT Photonic-Bands. To match the calculated edge state to the measured dispersion relation, the refractive index of silicon was taken as $n = 3.48$. To account for the corner roundness arising from fabrication, a fillet of $\SI{60}{nm}$ radius was added to the triangular holes of lattice constant $\SI{517}{nm}$. The unit cell consisted of equilateral triangles, with a larger triangle side length $d_1 = 0.6 a$ and a smaller triangle side length $d_2 = 0.3 a$.

\textbf{Device fabrication.} The PhC slab was fabricated on a silicon-on-insulator (SOI) platform with a $\SI{220}{nm}$ thick silicon layer on a $\SI{3}{\mu m}$ buried oxide layer. A positive electron-beam resist AR-P 6200.09 of thickness $\SI{240}{nm}$  was spin-coated. The PhC design was patterned using e-beam lithography on a Raith Voyager with \SI{100}{kV} beam exposure. The e-beam resist was developed in pentyl acetate/O-Xylene/isopropanol, and the chip underwent reactive-ion etching in HBr and Ar. In the next step, a photo-lithography resist S1813 was patterned using an EVG 620 NT mask aligner to define a wet-etching window around the PhC, not covering the ridge waveguides to prevent buckling. After development with MF321, the buried-oxide layer was removed in 7:1 buffered hydrofluoric acid for \SI{18}{mins}. The PhC was then cleaned in piranha solution and carefully removed from the solution, before being mounted in the near-field optical microscopy setup. 
The PhCs support an incoupling single TE-mode waveguide into the crystal to allow better index-matching for efficient coupling. \cite{shalaev2019robust}

Two different VPC domain walls were fabricated to facilitate transmission and back-scattering comparisons. The straight edge VPC has dimensions $215a \times 71a$, where $a = \SI{510}{nm}$ with the designed band gap falling within the excitation wavelength range $1480-\SI{1640}{nm}$. A total of 45 PhCs were measured, with 9 configurations of disorder. 

\textbf{Near-field optical microscopy setup.} 
The employed aperture-based near-field optical microscopy setup has been discussed in detail elsewhere \cite{Rotenberg2014}. In short, an aperture-based probe picks up the decaying evanescent fields on top of the VPC. It consists of a tapered optical fiber coated with $\SI{140}{nm}$ aluminum and an aperture of ca. $\SI{170}{nm}$ is created via focused ion beam milling. Scanning the probe over the silicon membrane at a relative height of ca. $\SI{20}{nm}$ is controlled via shear force. The amplitude and phase information are determined using a heterodyne detection scheme, with the coherent reference light beam shifted by $\Delta f = \SI{40}{kHz}$ in frequency.

\twocolumngrid
\bibliography{main}

\clearpage

%
%



\newcommand{\figref}[1]{Fig.~\ref{#1}}
\def\theequation{S\arabic{equation}}
\def\thefigure{S\arabic{figure}}
\def\thepage{S-\arabic{page}}



\onecolumngrid
\renewcommand{\thesubsection}{\Roman{subsection}}
\renewcommand\thefigure{S\arabic{figure}}    
\setcounter{figure}{0}    

%
\section*{Supplementary Information}


\subsection{Engineered disorder fabrication}

A total of 45 PhCs were fabricated, with $9$ configurations of disorder as 
\begin{equation}
    \sigma (0) = \text{Without intentional disorder},\\ 
\end{equation}
\begin{equation}
   \sigma (pos) = [ (0.0025,0), (0.005, 0), (0.01,0), (0.02,0)]a,\\
\end{equation}
\begin{equation}
     \sigma (size)  = [(0, 0.0025), (0, 0.005), (0, 0.01), (0,0.02)]L_1 or L_2
\end{equation}

where pos and size signify if the disorder induced was a position dislocation or size deformation. The position disorder scales with the lattice constant $a = \SI{510}{nm}$ and the size disorder scales with the relative hole size $L_1$ or $L_2$. For each degree of disorder in every bearded interface PhC, the random disorder is drawn using a normal (Gaussian) random number generator. To disregard large deviations in resist thickness, the 45 waveguides were fabricated in the central region of the SOI chip ($\SI{20}{mm} \times \SI{10}{mm}$), each separated by $\SI{100}{\mu m}$.

\begin{figure}[hbt]
	\centering
    \includegraphics[width=0.7\textwidth]{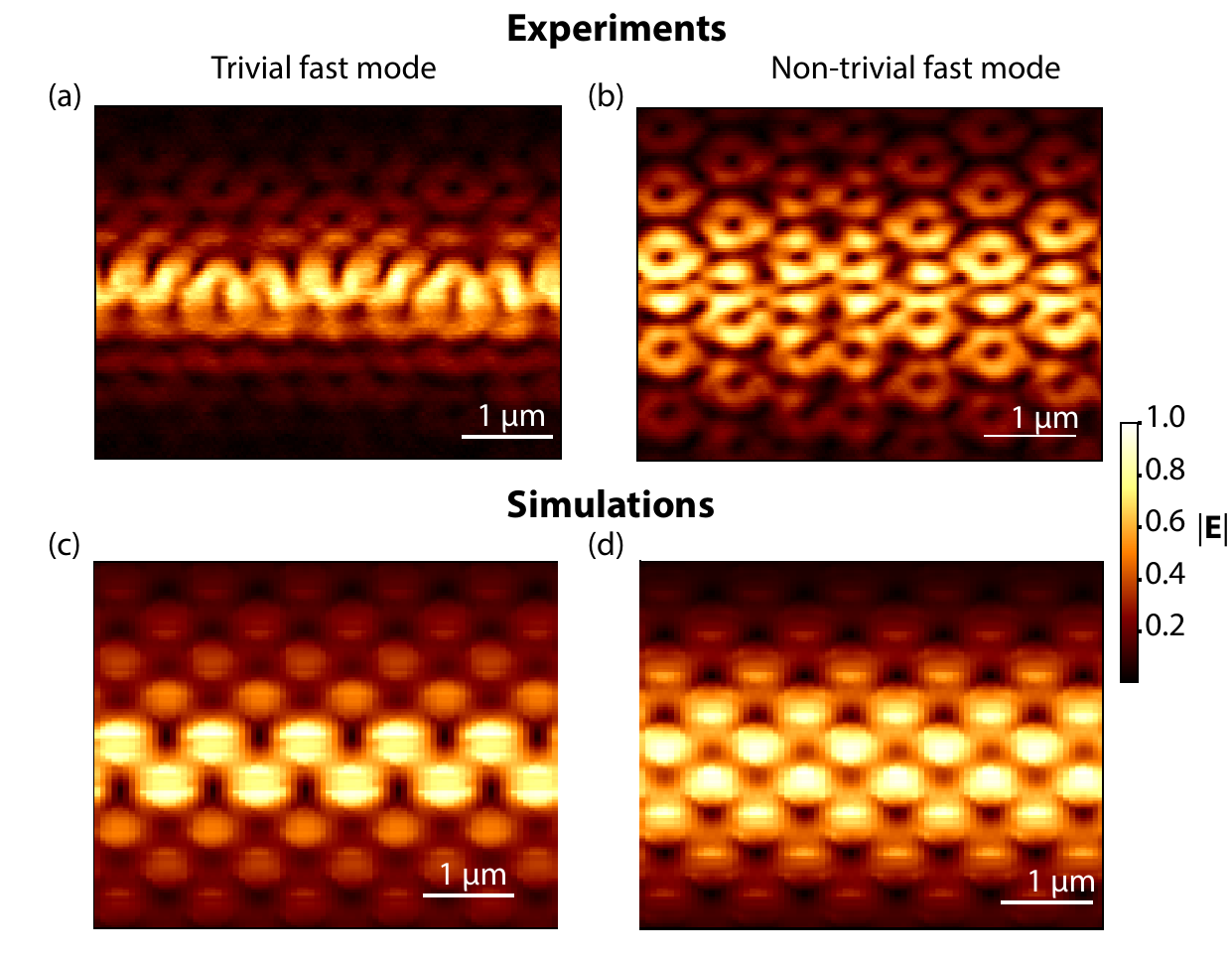}
	\caption{Electric field amplitude of (a) trivial and a (b) non-trivial fast mode of a pristine no disorder waveguide far away from the degeneracy point.Trivial mode at \SI{198.27}{THz} and nontrivial mode at \SI{189.74}{THz}. (c) and (d) show the corresponding numerically simulated electric field amplitude}
	\label{ch6supp:fig_electricfielmap}
\end{figure}
\subsection{Electric field amplitude}
\label{chap6_efield}

Fig. \ref{ch6supp:fig_electricfielmap}(a) and (b) show the experimentally measured in-plane field amplitude distribution for a trivial and non-trivial edge mode, respectively. It is important to note that the amplitude distribution shown here is from a different bearded interface PhC than discussed in the previous sections.  Fig. \ref{ch6supp:fig_electricfielmap}(a) shows a trivial fast edge mode at an excitation wavelength  $\lambda = \SI{1512}{nm}$ and a non-trivial fast mode at $\lambda = \SI{1580}{nm}$, both measured at an equidistant frequency from the degeneracy point at the edge of the Brillouin zone. For an illustrative $k$-point of $k_x=0.40 \cdot {2 \pi}/{a}$, where the trivial and non-trivial modes have a linear dispersion with a group velocity of $ \approx c/7$, we observe that the mode symmetries of the numerically determined eigenmode of the trivial and non-trivial edge mode match perfectly the experimentally measured mode profile.

\subsection{Group velocity fits and center wavelengths}
\label{chap6_groupvelocityfits}

\begin{figure}[h]
	\centering
    \includegraphics[width=\linewidth]{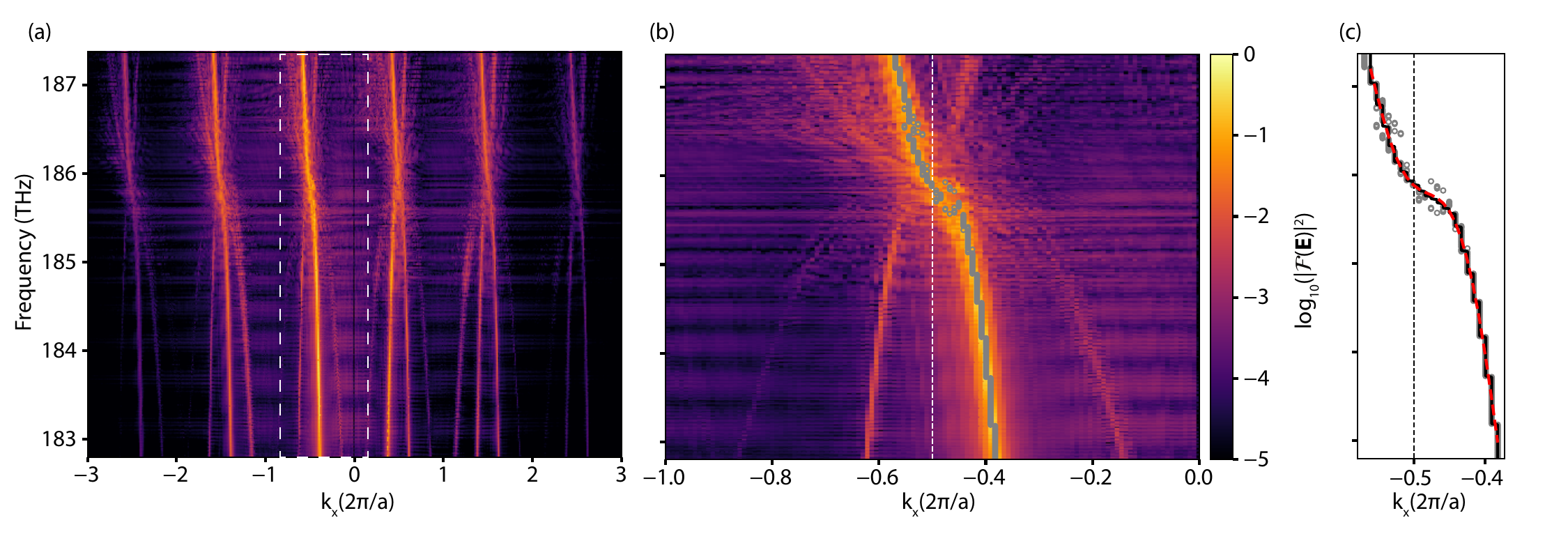}
	\caption{Dispersion and edge mode of an individual waveguide. (a) Experimentally retrieved dispersion of the edge mode. Bloch harmonics are separated by $2\pi/a$ with each a trivial and non-trivial edge mode degenerate at the Brillouin zone edge. The Fourier intensity is normalized to the maximum value and results in the $-\pi/a$ BH being the most prominent, as shown in (b). The grey points here denote the peak positions within this irreducible BZ. (c) The black line denotes the median filter applied to the peak positions. The red line denotes a polynomial fit of degree 6.}
	\label{ch6supp:fig1}
\end{figure}

An ensemble average without engineered random disorder, shown in Fig. 2 is built using $5$ individual bearded interface measurements. We use an exemplary individual interface, namely WG4 to discuss how the degeneracy points and group index curves were obtained. Using the near-field technique in collection mode, we raster-scan the probe over a length $ x = \SI{60}{\mu m}$ along the propagation direction and in the transverse direction $y = \SI{200}{nm}$.  We obtain a full dispersion relation of the edge mode by Fourier transforming the electric field intensity for each wavelength in the range $\lambda = \SI{1480}{nm} - \SI{1640}{nm}$ with a fine step wavelength resolution of $\SI{0.1}{nm}$, shown in Fig. \ref{ch6supp:fig1}(a). In the obtained spatial frequencies of the light, plotted on a logarithmic scale, we resolve at least $6$ higher-order Bloch harmonics (BH). We obtain a signal-to-background ratio of $\SI{21.36}{dB} \pm 0.9$. Since each higher-order BH carries a certain relative amplitude, we select the BH with the highest amplitude (in this case, it lies at $-\pi/a$). For each wavelength, we extract the corresponding wavevector to build the full mode shown as the overlaid grey circles in Fig. \ref{ch6supp:fig1}(b).  

\begin{figure}[hbt]
	\centering
    \includegraphics[width=\textwidth]{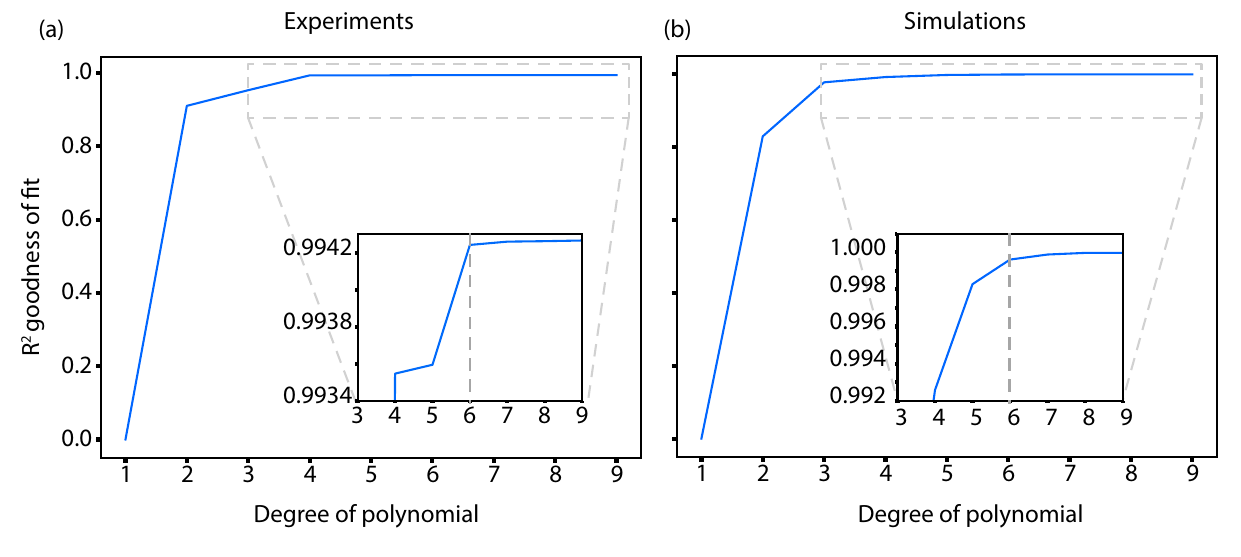}
	\caption{Calculated $R^2$ Goodness of fit for the polynomial fit of degrees $1 - 9$ from (a) experimental and (b) numerically simulated dispersion. Inset: zoom-in of the R-squared for the polynomial degree range $3-9$ to determine the best value for the degree of the polynomial for group velocity fits. The vertical dashed line for polynomial degree $6$ shows the chosen value.}
	\label{supp:goodnessfit}
\end{figure}

Due to the resolution in $k_x$ and an unclear dispersion around the degeneracy point, we find the median value of the peaks in the dispersion curve (black lines in Fig. \ref{ch6supp:fig1}). Since the non-trivial and trivial curves are not symmetric around the BZ edge, a higher degree polynomial is selected.  The choice of polynomial undergoes rigorous treatment before a decision is made. We use the coefficient of determination ($R^2$) as a metric to determine the goodness of the polynomial fit and therefore any discrepancy between observed and modeled values for the polynomial (see Fig. \ref{supp:goodnessfit}a). We select a hexic polynomial (degree-$6$) for which the calculated $R^2 = 0.995$ and the corresponding polynomial fit to the mode's wavevector is shown as a red curve in Fig. \ref{ch6supp:fig1}(c). A similar treatment was undergone to the simulations for clarity and confidence (see Fig. \ref{supp:goodnessfit}b).

\clearpage

\subsection{Degeneracy point determination}
\label{Supp-degeneracy}

\begin{figure}[hbt]
	\centering
    \includegraphics[width=\textwidth]{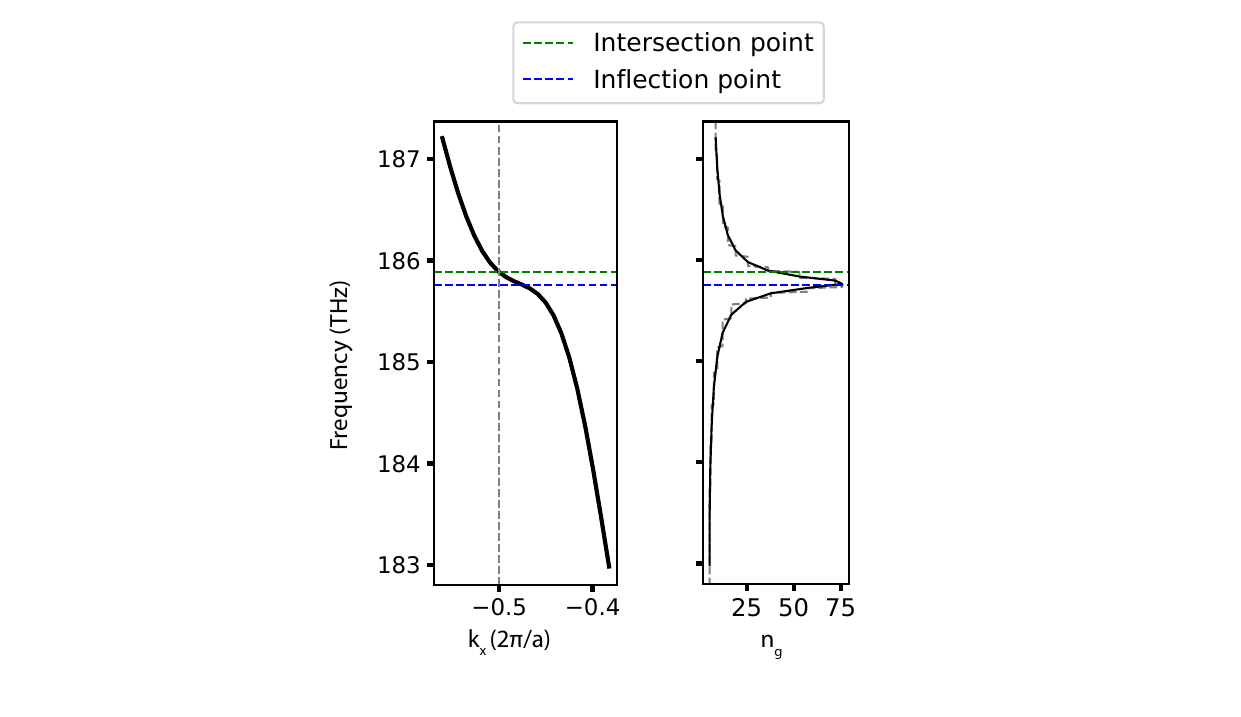}
	\caption{Determination of the degeneracy point. (a) 6-degree polynomial fit and the corresponding (b)  group index curve. The horizontal green dashed line indicates the intersection point of the fit to $k_x = 0.5 \cdot 2\pi/a$ and the horizontal blue dashed line denotes the inflection point of the fit.}
	\label{supp:DPchoice}
\end{figure}

The frequency at which the trivial and non-trivial modes become degenerate slightly shifts in the measured dispersion of individual waveguides due to fabrication imperfection. We identify the degeneracy point, as the intersection where the polynomial converges to the vertical line signifying the end of the Brillouin zone at $k_x=0.5 \cdot {2 \pi}/{a}$,  where $a = 0.510$ is determined from SEM images for the fabricated photonic crystals (see Fig. \ref{supp:DPchoice}). The inflection point of the fitted hexic polynomial assigns the wavelength where the sign of the fit undergoes a sign change, and therefore the wavelength at which maximum $n_g$ occurs. This tends to drastically vary for disordered photonic crystals that do not have a clear dispersion curve in the region where the slowdown factor is highest and depends solely on the polynomial fit at the fast light region. We observe an average difference of  $\SI{1.025}{nm} \pm \SI{0.46}{nm}$ between the inflection and intersection point, for 5 waveguides without the engineered disorder.

\newpage
\subsection{Subtleties in ${n}_g$ determination}
\label{Supp-ng_simualtions}
\begin{figure}[hbt]
	\centering
    \includegraphics[width=\linewidth]{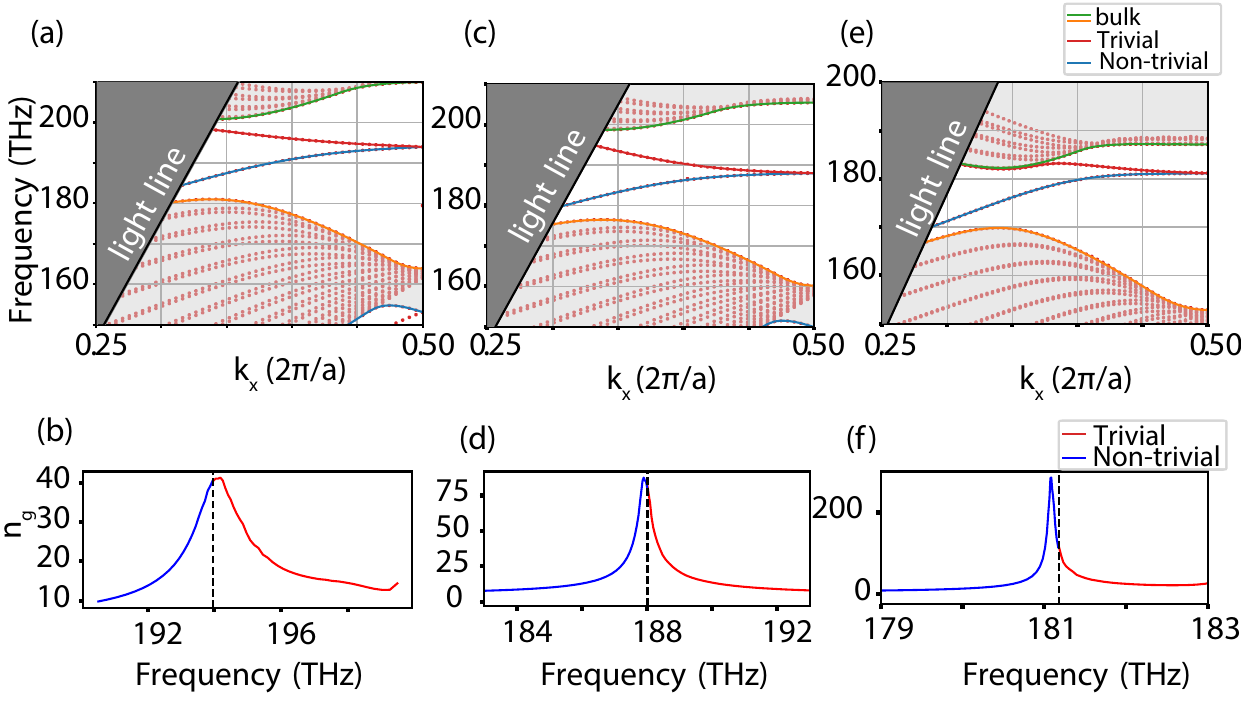}
	\caption{Group index curves in the slow light region. (a), (c), (e) Numerically simulated dispersion curves where the maximum group index lies in the (b) trivial region, (d) around the degeneracy point, and (f) non-trivial region, respectively. The parameters for the numerical simulation are $a = \SI{512}{nm}$, $L1 = 1.35a/\sqrt{3}= 0.78a$  and $L2 = 0.76a/\sqrt{3} = 0.44a$, (a) fillet radius $r = \SI{20}{nm}$ and (c) fillet radius $r = \SI{40}{nm}$, and (e) $a = \SI{508}{nm}$, $L1 = 1.04a/\sqrt{3}= 0.6a$  and $L2 = 0.7a/\sqrt{3} = 0.4a$ and , fillet radius $r = \SI{1}{nm}$. }
	\label{supp:maxngregion}
\end{figure}

Slight variations in the geometry such as rounding of the corners and the relative sizes of the holes result in changes in the dispersion and therefore variations in the $n_g$ curves as a function of frequency. We find that therefore, the frequency range where the maximum group index occurs can be affected crucially. In Fig. \ref{supp:maxngregion}, we show three numerically simulated dispersion curves and their corresponding group index curves. The parameters used for Fig. \ref{supp:maxngregion}(a) and (b) were $a =\SI{512}{nm}$, $L1 = 1.35a/\sqrt{3}= 0.78a$  and $L2 = 0.76a/\sqrt{3} = 0.44a$ and fillet radius of $\SI{20}{nm}$. The maximum $n_g = 41 $ occurs in the trivial region given by the red curve. This coincides well with the observation reported in ref. \cite{rosiek2022observation}. On increasing the roundedness to a fillet radius of \SI{40}{nm} of the triangular holes, while ensuring that the edge modes remain single-mode and do not exhibit intermodal scattering, we observe that the DP shifts to lower frequencies and the edge modes become degenerate around $\SI{188}{THz}$ (see Fig \ref{supp:maxngregion}(b)). The corresponding group index results in a maximum value very close to the DP with $n_g = 87$. In Fig. \ref{supp:maxngregion}(c), the numerically simulated dispersion for a lattice constant $a=\SI{508}{nm}$ and $L1 = 0.6a$ and $L2 = 0.4a$ show a further reduction in the DP to lower frequencies and the corresponding group index curve features a maximum $n_g = 285$ in the topological region, consistent with the observation in ref. \cite{yoshimi2020slow} and ref. \cite{Yoshimi2021}. It is clear that the relative hole sizes, and the rounding of the triangles due to fabrication variations can consistently shift the maximum $n_g$ obtained and must be taken into account for experimentally realizing topological edge states in the slow-light region.

\newpage
\subsection{Spatial map }
\begin{figure}[hbt!]
	\centering
    \includegraphics[width=0.7\textwidth]{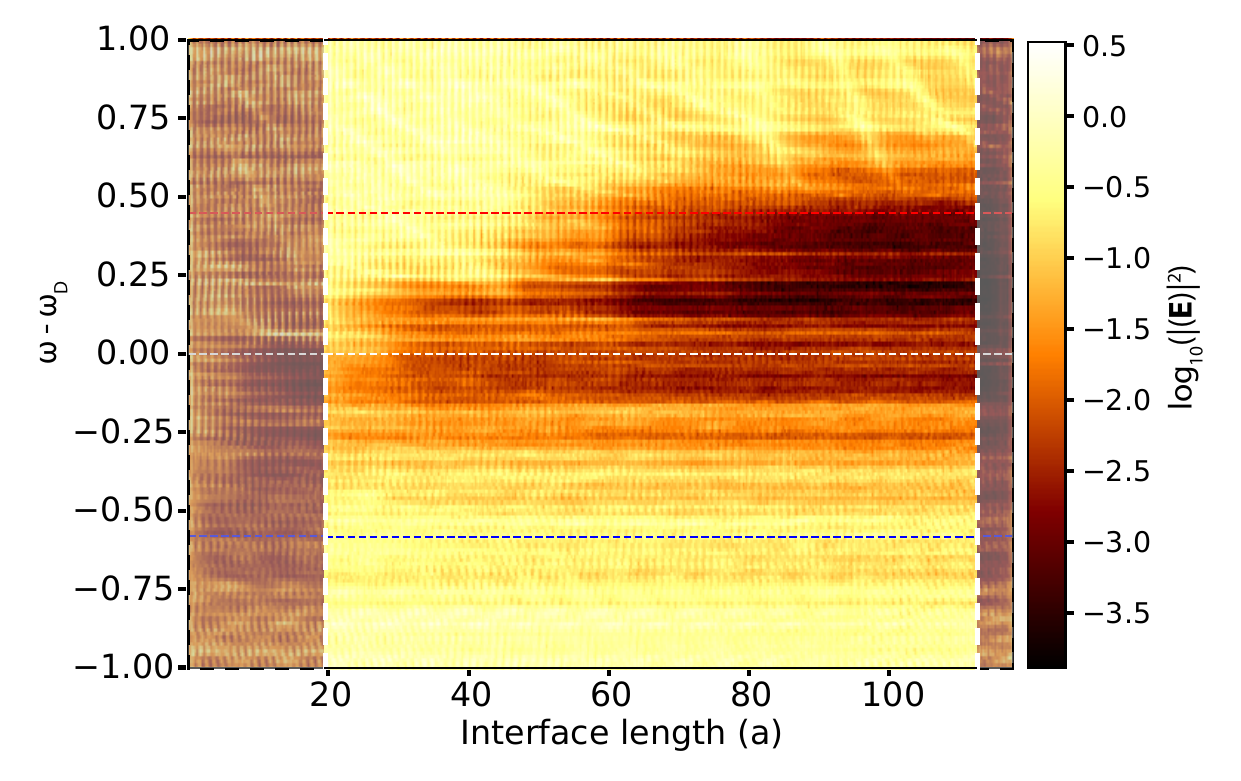}
	\caption{Slow light spatial-spectral map. A cropped normalized intensity map on a logarithmic scale is overlaid on top of the spatial-spectral amplitude map shown in Fig. 6a where the cropped interface length is in the range $[19.61 - 112.64]$ in units of $a$.}
	\label{supp:spatialslow}
\end{figure}


\subsection{Calculated BMFP and mean group indices }

\begin{table}[h!]
    \centering
    \begin{tabular}{ |p{1.5cm}||p{1.5cm}|p{1.5cm}|p{1.5cm}|p{1.5cm}| |p{3.2cm}|} 

 \hline
 Disorder magnitude &  Non-trivial BMFP & Trivial BMFP & Non-trivial $\overline{n_g}$ & Trivial $\overline{n_g}$ & \Large $\frac{|\Delta n_g|}{\frac{1}{2}(\overline{n_{g,T}} + \overline{n_{g,NT}})} (\%)$ \\ [0.5ex] 
\hline\hline
0 & 3861.44 & 636.86 & 10.60 & 10.71 & 1.05 \\ 
\hline\hline
\multicolumn{6}{|c|}{Position} \\
\hline
0.25\% & 1879.33 & 275.79 & 9.95 & 8.90 & 11.17 \\ 
0.5\% & 1074.44 & 440.83 & 7.42 & 8.29 & 11.17 \\ 
1 \% & 520.26 & 313.88 & 8.21 & 7.83 & 4.75 \\ 
2 \% & 532.52 & 467.98 & 11.56 & 11.19 & 3.28 \\ 
\hline\hline
\multicolumn{6}{|c|}{Size} \\
\hline
0.25\% & 2501.02 & 575.71 & 8.94 & 9.17 & 8.27 \\ 
0.5\% & 2231.78 & 400.25 & 10.19 & 8.91 & 13.39 \\ 
1 \% & 2346.43 & 334.34 & 9.04 & 9.61 & 6.07\\ 
2 \% & 680.49 & 591.04 & 7.68 & 6.97 & 9.70 \\ 
\hline
\end{tabular}
    \caption{Overview of ensemble-averaged calculated BMFPs and group indices for 9 magnitudes of disorders. The columns denote (1) the magnitude of the engineered disorder (position or size), calculated BMFP for the fast (2) non-trivial (NT) and (3) trivial (T) region, mean group index of the fast light range for  (4) non-trivial and (5) trivial region, and the deviation of mean group indices.}
    \label{supp_BMFP_ngtable}
\end{table}
\begin{center}
\end{center}


\end{document}